# When is an article actually published? An analysis of online availability, publication, and indexation dates

Stefanie Haustein[1], Timothy D. Bowman[1] and Rodrigo Costas[2]

[1] *stefanie.haustein@umontreal.ca, timothy.bowman@umontreal.ca*
École de bibliothéconomie et des sciences de l'information, Université de Montréal, Montréal (Canada)

[2] *rcostas@cwts.leidenuniv.nl*
Center for Science and Technology Studies, Leiden University, Wassenaarseweg 62A, 2333 AL Leiden (The Netherlands)

**Abstract**
With the acceleration of scholarly communication in the digital era, the publication year is no longer a sufficient level of time aggregation for bibliometric and social media indicators. Papers are increasingly cited before they have been officially published in a journal issue and mentioned on Twitter within days of online availability. In order to find a suitable proxy for the day of online publication allowing for the computation of more accurate benchmarks and fine-grained citation and social media event windows, various dates are compared for a set of 58,896 papers published by Nature Publishing Group, PLOS, Springer and Wiley-Blackwell in 2012. Dates include the online date provided by the publishers, the month of the journal issue, the Web of Science indexing date, the date of the first tweet mentioning the paper as well as the Altmetric.com publication and first seen dates. Comparing these dates, the analysis reveals that large differences exist between publishers, leading to the conclusion that more transparency and standardization is needed in the reporting of publication dates. The date on which the fixed journal article (Version of Record) is first made available on the publisher's website is proposed as a consistent definition of the online date.

**Conference Topic**
Journals, databases and electronic publications

**Introduction**
The process of scholarly communication, which usually begins with the formulation of a research idea and hypothesis and ends with publishing results to share them with the scientific community (Garvey & Griffith, 1964), has been sped up by means of electronic publishing (Dong, Loh, & Mondry, 2006; Wills & Wills, 1996). The publication delay, which Amat (2008, p. 382) defined as the "chronological distance between the stated date of reception of a manuscript by a given journal and its appearance on any print issue of that journal", has been accelerated by email and online manuscript handling systems as well as online publication (Wills & Wills, 1996). The delay period consists of the review process, which constitutes the main delay and ends with the acceptance of the manuscript, followed by technical delays of journal production and paper backlog.

Various studies have analyzed publication delays and found differences between scientific fields, journals, and publishers (e.g., Abt, 1992; Amat, 2008; Björk & Solomon, 2013; Das & Das, 2006; Diospatonyi, Horvai, & Braun, 2001; Dong et al., 2006). Since long delays interfere with priority claims and slow down scientific discourse, publication speed plays an important role for authors and scholarly communication (Rowlands & Nicholas, 2006; Schauder, 1994; Tenopir & King, 2000). Short publication delays can therefore be considered as a quality indicator reflecting the up-to-dateness of scientific journals (Haustein, 2012). Publishers have begun to reduce delays by making so-called *early view*, *in press*, *ahead of print* or *online first* versions of accepted papers available before they appear in an (print) issue. It has been shown for food research journals that online ahead of print publication has reduced publication delay by 29% (Amat, 2008), while Das and Das (2006) reported for 127 journals in 2005 average lags of three months between online and print issues publications with particular differences between publishers. Tort, Targino, and Amaral (2012) showed that this lag increased significantly over time for six neuroscience journals. Online dates are now being recorded in bibliometric databases like Scopus, which impacts bibliometric analyses (Gorraiz, Gumpenberger, & Schlögl, 2014; Heneberg, 2013). Together with the increasing popularity of preprint servers (such as arXiv and SSRN) and institutional repositories, such *in press* versions have helped to speed up the read-cite-read cycle. As a result manuscripts increasingly cite papers that have not been officially published in a journal issue. Although scholarly communication has always involved sharing different versions of a manuscript with colleagues before, during, and after formal publication—such as exchanging drafts for feedback before submission or diffusing preprints after acceptance—, the electronic era makes these versions 'public', searchable, and (often) permanently retrievable on the web. To define and distinguish between various versions, the National Information Standards Organization (NISO) agreed upon the following versions of a journal article (NISO/ALPSP Working Group, 2008):

- Author's Original (AO) – manuscript ready to submit.
- Submitted Version Under Review (SMUR) – manuscript under formal peer review.
- Accepted Manuscript (AM) – version of journal article accepted for publication.
- Proof (P) – copy-edited version of accepted article.
- Version of Record (VoR) – fixed version of journal article formally published.
- Corrected Version of Record (CVoR) – VoR in which errors have been corrected.
- Enhanced Version of Record (EVoR) – VoR updated or enhanced with supplementary material.

It is important to note that by the NISO definition, the VoR is defined as a "fixed version of a journal article that has been made available by any organization that acts as a publisher by formally and exclusively declaring the article 'published'" (NISO/ALPSP Working Group, 2008, p. 3). This definition includes early views and in press articles without information on volume and issue or other identifiers as long as the content and layout of the article are fixed.

When it comes to bibliometric indicators, the acceleration of the publication process has been reflected in obsolescence patterns (Egghe & Rousseau, 2000) as well as citing half-lives (Luwel & Moed, 1998). These increasing online-to-print lags were shown to artificially increase citation rates including the immediacy index and impact factor (Heneberg, 2013; Seglen, 1997; Tort et al., 2012). The speed of scholarly communication becomes particularly visible in the context of social media metrics (the so-called altmetrics); for example, mentions of scientific documents on Twitter happen within hours (and sometimes within minutes) of online availability (Shuai, Pepe, & Bollen, 2012).

We argue that in the fast-moving digital era, the use of the publication *year* of the journal issue as the smallest level of time aggregation for bibliometric indicators is becoming insufficient, particularly in research evaluation contexts, due to the following factors:

a. acceleration of the read-cite-read cycle due to electronic publishing,
b. commonplace of online publication before publication of the journal issue, and
c. increasing online-to-print lags.

Following NISO's terminology, we suggest that the date of the first public online appearance of the VoR is the most relevant and should be used as the basic time unit to determine the official publication date of a paper. This would allow for the construction of more accurate citation and social media event windows, for example, citation windows of equal length (in days or months) for papers published in January or December, as well as the construction of more exact benchmarks by aggregating citations and social media events per week (e.g., tweets and Facebook shares) or month (citation rates) depending on the evaluation context.

Although many publishers now report online publication dates, many different dates are presented and the information provided varies between publishers, as no official standards exist on publication dates. This paper explores and aims to verify various 'publication' dates in order to find a good proxy for the actual date of online availability. Thus, the paper aims to answer the following research questions:

1. Which publishers specify online dates and how do they provide them?
2. How reliable are dates provided by the publishers and how do they compare to each other?
3. What other existing dates can be used as a proxy of the online publication date of the VoR?

**Methods and Materials**

The dataset of this study was retrieved from the Web of Science (WoS) (as the major citation database) and is restricted to the publication year 2012 to limit effects of changes over time. To validate the publication dates provided by the publishers, the dates of the first tweet mentioning the particular paper were obtained from Altmetric.com. We argue that a tweet cannot link to a paper before it exists, thus the first tweet cannot have appeared before the online publication date. Tweets captured by Altmetric.com are linked to the documents via the DOI resulting in 313,301 WoS 2012 papers with at least one event captured by Altmetric.com (Haustein, Costas, & Larivière, 2015). Altmetric records that contained an arXiv ID or Astrophysics Data System (ADS) ID were removed to exclude tweets to preprints, which could have been made public before the online publication of the VoR. Twitter mentions are thus restricted to the mentions or links to the publisher's website, DOI, or PubMed ID.

Table 1. Top 10 publishers according to number of papers with types of dates available according to data provided by the publisher via API (a), in the metadata (m) of the webpage, on the webpage only (w), or as dynamic content only (d). Publishers selected for this study are highlighted in grey.

| Publisher | Papers | Received | Revised | Accepted | Version of Record | Online | Publication | Date | Journal Issue | Journal Issue Online |
|---|---|---|---|---|---|---|---|---|---|---|
| Elsevier | 51,292 | d | d | d | | d | a | | w | |
| Wiley-Blackwell | 47,958 | w | | w | | m,w[i] | m | | w,m | w |
| Lippincott | 21,944 | | | | | | | m | w,m | |
| Springer | 19,225 | | | | | m | m,a | m | w,m,a | |
| PLOS | 16,208 | w | | w | | | a,m | | a,m | |
| BMC | 11,930 | w | | w | | | w,m | | w,m | |
| NPG | 11,181 | w,m | | w,m | | m,a | w,m,a | | w,m,a | |
| ACS | 11,024 | | | | | | | m,w | w | |
| Oxford | 10,368 | w | | w | | w | | m | w,m | |
| Sage | 8,776 | | | w | | w | | m | w,m | |

[i] Wiley provides two online dates "article published online" as well as "online date". See explanations below.

The top 10 publishers[1] of papers in the WoS-Altmetric dataset can be found in Table 1 together with the date information provided via API, in the metadata, in the webpage only, or as dynamic content of the webpage. It can be seen (in the headings of the table) that multiple terms exist to describe the online publication date and that multiple types of dates are made available on the website, in the metadata, or via the API; these include received, revised, accepted, version of record, online, publication, and date. Based on checking samples of articles for each of the publishers, we assume that the dates provided as *Version of Record*, *Online*, *Publication* and *Date* (Table 1) refer to (first) online appearances of the VoR required for this study. Wiley-Blackwell, Springer, PLOS, and Nature Publishing Group (NPG) were chosen due to their coverage and the technical feasibility of retrieving online date information. While Elsevier was the most represented publisher in this sample, it was difficult to obtain the required date information for their articles using PHP because this information is inserted dynamically into the webpage using JavaScript; Elsevier offers an API, but when queried[2] it was found to provide access to only the issue date and not to the online publication dates required for this study.

Using the DOI, the respective publishers' web platforms were queried to retrieve online dates. PLOS, Springer, and NPG each offer an API, but it was found that in some instances additional date information was only made available by searching the web page. In order to obtain the dates for Wiley, Springer and NPG, a PHP script was written that retrieved the HTML of the page. The HTML was then searched for metadata containing date information (e.g. `<meta name="prism.publicationDate" content="2012-01-05"/>`). When date information was found, it was saved to a relational database for evaluation. In instances where the article website had no (or missing) metadata available, the HTML was parsed and the contents of specific HTML tags found to contain date information was extracted and saved to a relational database; for the Wiley articles, a second script was written to retrieve dates not found in the metadata.

To compare different dates available and test in how far they can be used as proxies for online publication dates, other date information was obtained from WoS and Altmetric, so that together with the information from publishers the following dates were available:
- *online date*: retrieved from the publishers websites as part of the article metadata. For NPG ("Advance Online Publication"[3]), Springer ("Online First"[4]), and Wiley-Blackwell ("Early View"[5]) this date marks when the VoR was made publically available on the publisher's website. For PLOS the online date equals the publication date because there is no difference between online and issue dates.
- *journal issue date*: the date from the journal issue as recorded by WoS. Since only a minority of papers provided the day of the month, the journal issue date was converted to the first of each month. Based on all 1.3 million papers in WoS published in 2012, 3.2% were published in issues spanning several months (such as JAN-FEB for a double issue). These were converted to the first day of the first month. A small percentage (0.5%) of papers appeared in seasonal issues (SPR, SUM, FAL, WIN). Since the data indicates that these are published at the beginning, middle, as well as the end of the particular season, these dates were disregarded. An additional 11.3% of all 2012 papers did not provide any issue date. Figure 1 provides an overview of the distribution of the 1.3 million WoS 2012 papers per journal issue date information.

---

[1] Publisher names from WoS were cleaned searching for name variants, but mergers and acquisitions were not accounted for. For example, BMC is considered and independent publisher, although it was acquired by Springer in 2008.
[2] Using the http://api.elsevier.com/content/abstract/doi/{doi} API call
[3] http://www.nature.com/authors/author_resources/about_aop.html
[4] http://www.springer.com/authors/journal+authors/helpdesk?SGWID=0-1723213-12-817311-0
[5] http://olabout.wiley.com/WileyCDA/Section/id-404512.html#ev

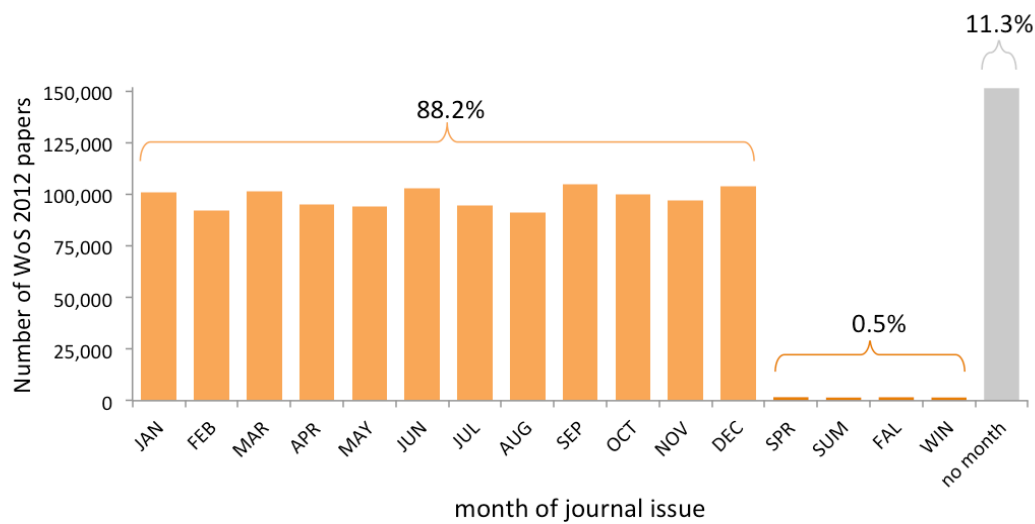

Figure 1. Number of WoS 2012 papers per months of journal issue.

- *Altmetric publication date*: the publication date as recorded by Altmetric.com, which is a mix of the journal issue date and online date (personal communication with Euan Adie and Jean Liu) as retrieved from the publisher. This is also the date Altmetric.com uses to compute the Altmetric score and provide benchmarks for papers of the same age. As shown in Figure 2, particular peaks can be observed for January 1 of each year as well as the first or last of each month. This might reflect common publishing practices, but could also be caused by aggregating data without actual day (and month) information. It was found that 15.1% of Altmetric.com records[6] did not have any publication date or they had incorrect dates (e.g. dates up to 2037).

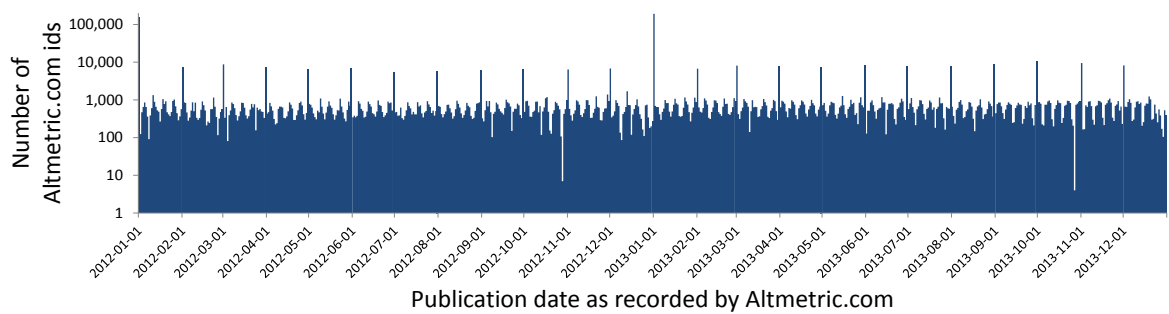

Figure 2. Number of Altmetric.com ids per Altmetric.com publication date from January 2013 to December 2014.

- *Altmetric first seen date*: the datestamp when Altmetric.com captured the first event for a particular document, which is missing for 4% of all records[7].
- *First tweet date*: the datestamp of the first tweet[8] captured by Altmetric.com (excluding all papers with links to arXiv IDs or ADS IDs to ensure that the tweet did not refer to a preprint).

---

[6] Based on 2.1 million Altmetric.com records collected in August 2014.
[7] Based on 2.1 million Altmetric.com records collected in August 2014.

- *WoS indexing date*: the day when the document was indexed by WoS, which for 2012 papers was mostly during (37.7%) or in the month before (11.5%) or after (29.4%) the journal issue month.

In addition to the dates above we were also able to retrieve the following information for the papers published by Wiley-Blackwell:
- *Manuscript received*: the date the AO was submitted.
- *Manuscript accepted*: the date the AM was accepted.
- *Article first published online*: we could not determine the exact meaning of this date; for 95.6% of the total 34,507 Wiley-Blackwell documents it was identical with the *online date* and for 1.6% it was missing. For 2.3% of papers the *article first published online date* occurred before the *online date* by, on average, 35 days, which suggest that it marks the publication of the AM. However, in 137 cases (0.4%), it followed the *online date* by, on average, 52 days.

The final dataset—that is, the match of WoS, Altmetric.com, and papers with online dates retrieved from the four publishers—included 71,175 papers. For better comparison, it was restricted to papers for which all five dates tested as proxies for online publication (i.e., journal issue, Altmetric publication and first seen date, first tweet and WoS indexing date) were available. This amounted to a total of 58,896 papers, 12.5% NPG, 16.3% PLOS, 24.6% Springer and 46.6% Wiley-Blackwell.

**Results and Discussion**

Descriptive statistics comparing the online date to the five potential proxies are presented in Table 2, highlighting particular differences for the four publishers. Based on the assumption that the online date provided by the publishers were correct, the Altmetric publication date, first seen date, as well as the first tweet date seem to be the best proxies for online publication, while the journal issue and WoS indexing date show the largest deviations from the online publication dates. These differences reflect the nature of these dates. For example, Altmetric collects its publication dates from the publishers websites and while first tweets are known to happen shortly after publication (Shuai et al., 2012), WoS processing takes more time, namely, on average between 39 days for PLOS or 163 days for Springer papers. The 61 (NPG), 84 (Wiley-Blackwell), and 146 (Springer) days between online and journal issue date mostly reflect the backlog between online availability and publication of the journal issue. Although the (print) issue is generally assumed to follow online publication chronologically, results in Table 2 show that for 3.47% of Springer, 9.09% of Wiley-Blackwell, and 20.04% of NPG papers analyzed the online date came after the journal issue date, which is considered negative delay (Das & Das, 2006).

Although Altmetric and Twitter dates work better than journal issue and WoS indexing, none of the dates seem to reflect the online date well and large differences can be observed between publishers, in particular for Wiley-Blackwell, which questions the validity of any of the five dates as a reliable proxy of the publication of the VoR across publishers. The Altmetric publication date, which overall shows the smallest difference compared to the online date provided by the publishers—on average, 9 days for Springer, 12 days for NPG, 27 days for PLOS, and 121 for Wiley-Blackwell—is also problematic, because it is set to a date prior to online publication in 43.37% of Springer, 55.38% of NPG, 63.83% of Wiley-Blackwell, and

---

[8] Twitter is the most common source covered by Altmetric.com (Robinson-García, Torres-Salinas, Zahedi, & Costas, 2014), so it makes sense to work with this date and not from other less common sources (e.g. Facebook or blogs).

66.49% of PLOS papers. The variance between publishers affects Altmetric scores (but arguably also citation scores) when benchmarking a paper's scores against that of papers of the same reported age.

**Table 2. Statistics for chronological distance (in number of days) of the journal issue month, Altmetric publication and first seen date, first tweet date and WoS indexing date with the online date for NPG, PLOS, Springer and Wiley-Blackwell.**

| Chronological distance to online date in number of days | | NPG *n=7,391* | PLOS *n=9,600* | Springer *n=14,473* | Wiley-Blackwell *n=27,432* |
|---|---|---|---|---|---|
| **Journal issue month**[i] | % before | 20.04% | | 3.47% | 9.09% |
| | % identical | 5.47% | | 0.11% | 0.29% |
| | % after | 74.50% | n/a[ii] | 96.42% | 90.62% |
| | mean | 61 | | 146 | 84 |
| | standard deviation | 78 | | 111 | 93 |
| | min | -330 | | -269 | -423 |
| | max | 548 | | 1,850 | 1,032 |
| **Altmetric publication date** | % before | 55.38% | 66.49% | 43.37% | 63.83% |
| | % identical | 39.35% | 31.41% | 34.11% | 2.81% |
| | % after | 5.28% | 4.44% | 22.52% | 33.36% |
| | mean | 12 | 27 | 9 | 121 |
| | standard deviation | 68 | 79 | 48 | 322 |
| | min | -3,013 | -697 | -519 | -16,761 |
| | max | 411 | 526 | 1,850 | 5,016 |
| **Altmetric first seen date** | % before | 3.48% | 0.00% | 0.08% | 14.59% |
| | % identical | 32.88% | 36.64% | 1.04% | 14.26% |
| | % after | 63.64% | 63.36% | 98.89% | 71.15% |
| | mean | 35 | 12 | 90 | 63 |
| | standard deviation | 87 | 49 | 164 | 122 |
| | min | -459 | 0 | -257 | -533 |
| | max | 890 | 602 | 1,843 | 1,228 |
| **First tweet date** | % before | 3.52% | 0.00% | 0.08% | 14.52% |
| | % identical | 34.37% | 37.23% | 1.06% | 15.21% |
| | % after | 62.21% | 62.77% | 98.85% | 70.27% |
| | mean | 37 | 15 | 92 | 65 |
| | standard deviation | 92 | 59 | 169 | 127 |
| | min | -459 | 0 | -257 | -533 |
| | max | 890 | 811 | 1,843 | 1,393 |
| **WoS indexing date** | % before | 2.72% | 0.00% | 0.10% | 0.05% |
| | % identical | 0.01% | 0.00% | 0.00% | 0.00% |
| | % after | 97.27% | 100.00% | 99.90% | 99.95% |
| | mean | 83 | 39 | 163 | 97 |
| | standard deviation | 81 | 20 | 113 | 94 |
| | min | -302 | 9 | -252 | -359 |
| | max | 576 | 262 | 1,866 | 1,049 |

[i] First of the journal issue month as recorded by WoS.
[ii] PLOS does not distinguish between online and issue date, so that the two dates are actually identical.

Based on the assumption that a tweet cannot mention a paper before it exists in the online space it links to, the online dates provided by Wiley-Blackwell seem to be the most problematic (Figure 3), as 14.52%[9] of the 27,432 analyzed papers had tweets linking to them before the date that the publisher identifies as the online publication date. On the other hand, none of the PLOS papers and few of the Springer (0.08%) articles were mentioned on Twitter before the online publication date. Although all of the papers analyzed have been tweeted, the mean number of days between online date and first tweet was higher than expected, ranging from 15 days for PLOS to 92 days for Springer. Moreover, the first mention on Twitter happened on the day of online publication for 1.06% (Springer) and 34.47% (NPG) sampled papers, which—particularly considering that about 80% of recent papers are never tweeted (Haustein, Costas, & Larivière, 2015)—limits the usefulness of the first tweet date as a proxy for online publication.

---

[9] Results change only slightly when using the *article first published online* date, i.e. 14.61% of Wiley-Blackwell papers had a tweet appear before this date.

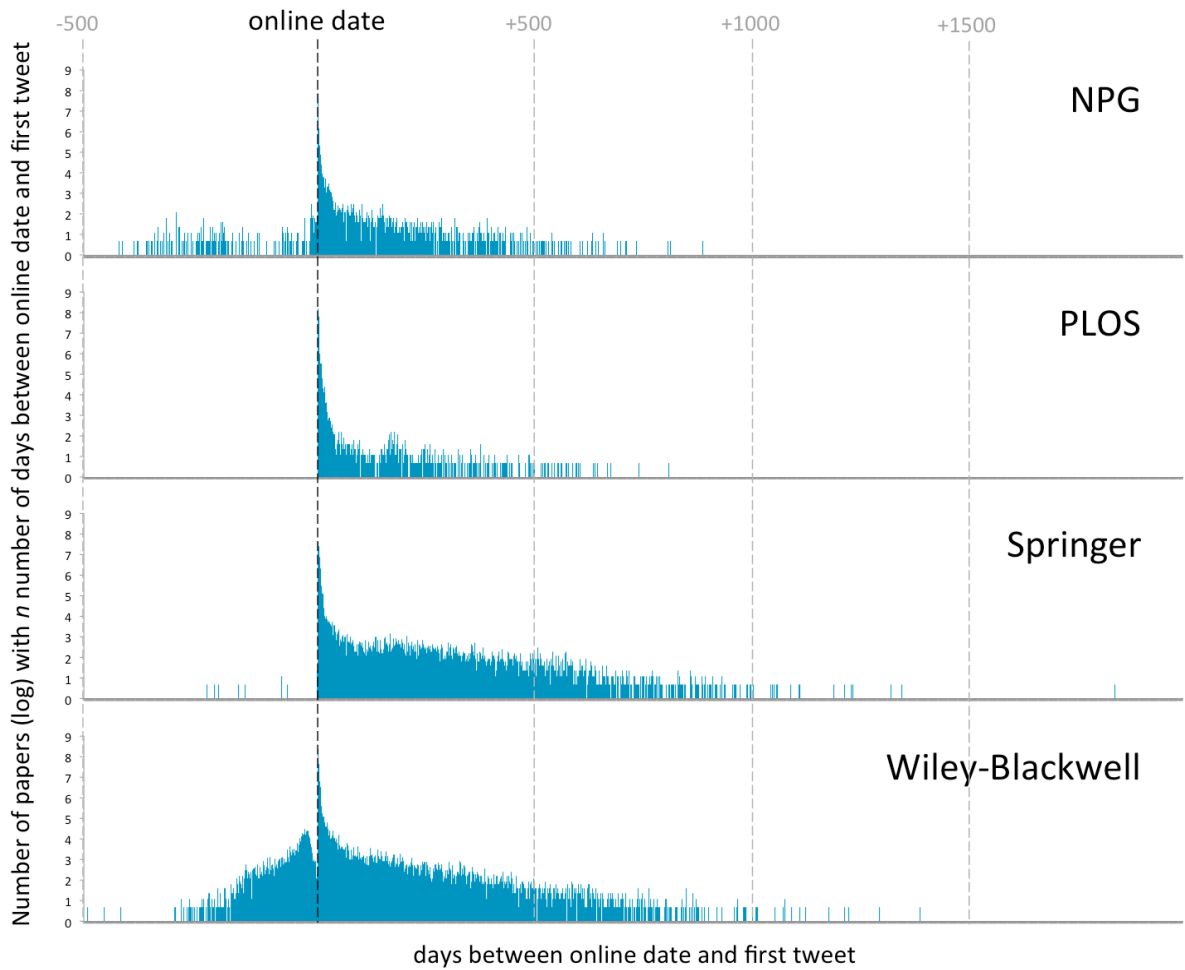

**Figure 3. Number of papers (log) with *n* days between online date and first tweet per publisher.**

**Conclusions and Outlook**

Currently none of the investigated dates represent a good proxy for the date a journal article was actually available online. In particular, the finding that a considerable amount of Wiley-Blackwell papers had been mentioned on Twitter before the online date, suggests that inconsistencies exist in terms of how publishers report online dates. This applies to the technical aspects as well as to actual content and vocabulary used. Thus, even when online dates can be retrieved from the publishers' websites or via API, they do not seem to always (and in a similar way for every publisher) mark the actual point in time when something was made accessible online. There is, thus, an urgent need for transparency and standardization of various dates reported by publishers in order to assure comparability of online dates across publishers. Adopting the vocabulary developed by NISO, specific dates could be reported for each version of the journal article, and the first appearance of the VoR would thus mark the date the fixed version of the document appeared online. A standardized vocabulary and a common definition of what various publication dates mean would not only improve benchmarking in the context of research evaluation but would also help to accurately determine the start of open access embargo periods required by certain funders, such as the NIH in the United States or the European Research Council. Currently these embargo periods, delaying green open access by a couple of months to years to protect publishers' revenue, are supposed to begin with publication of the article, which can refer to either journal issue or online date.[10] Setting the start date of the embargo to the online publication date of the VoR would remove a potential loophole that allows the publishers to increase the embargo period during which they have the exclusivity of access.

Until such a standard is implemented, research on metrics should focus on obtaining more publisher-independent date information. One potential proxy for online publication could be the date when a DOI resolved successfully for the first time. Recently CrossRef has implemented the DOI Chronograph, a tool which tracks various deposits of metadata by the publisher as well as the first day of successful DOI resolution (Wass, 2015). Future work will investigate in how far these dates can be used to create fine-grained benchmarks needed in the context of social media metrics. Regarding citations, where monthly proxies are sufficient, the WoS Indexing date should be further investigated.

**Acknowledgments**

The authors would like to thank Euan Adie and Altmetric.com for access to their data and acknowledge funding from the Alfred P. Sloan Foundation, grant no. 2014-3-25.

---

[10] http://authorservices.wiley.com/bauthor/faqs_fundingbodyrequirements.asp